\documentstyle{mn}

\newcommand{\R}{ROSAT }
\newcommand{\A}{ASCA }
\newcommand{\src}{GS\,0834--43} 
\newcommand{\etal}{et al.}

\def\ergs{\rm erg\,s^{-1}}

\def\ltsima{$\; \buildrel < \over \sim \;$}
\def\lsim{\lower.5ex\hbox{\ltsima}} 
\input psfig.tex 
\newcommand{\rc}{\rm}
\title[The optical/IR counterpart of the 12s transient X--ray pulsar \src]
{The discovery of the optical/IR counterpart of the 12\,s transient X--ray pulsar 
\src\,\thanks{Partially based on observations carried out at ESO, La Silla, Chile 
(62.H--0513)}}
\author[Israel et al.]{G. L. Israel,$^{1,}$\thanks{Affiliated to ICRA.}
S. Covino,$^2$ S. Campana,$^{2, \dag~}$ V.F. Polcaro,$^3$ P. Roche,$^4$   
L. Stella,$^{1,\dag}$ \newauthor A. Di Paola,$^1$ 
D. Lazzati,$^{2,5}$ S. Mereghetti,$^6$ E. Giallongo,$^1$ A. Fontana,$^1$ 
\newauthor and 
F. Verrecchia$^7$ 
\\ 
$^1$Osservatorio Astronomico di Roma, Via Frascati 33, 
I--00040 Monteporzio Catone (Roma), 
Italy\\ 
$^2$Osservatorio Astronomico di Brera, Via E. Bianchi 46, 
I--23807  Merate (Lecco), Italy\\ 
$^3$ Istituto di Astrofisica Spaziale, Area di Ricerca di Roma--Tor Vergata, 
Via Fosso del Cavaliere, I--00133 Roma, Italy\\  
$^4$ Dept. of Physics \& Astronomy, Univ. of Leicester, Leicester, UK\\ 
$^5$ Dipartimento di Fisica, Universit\`a degli Studi,
via Celoria 13,  I--20133 Milano, Italy\\
$^6$ Istituto di Fisica Cosmica ``G.P.S. Occhialini'' del C.N.R., 
Via Bassini 15, I--20133 Milano, Italy\\
$^7$ Universit\`a ``La Sapienza'', Dipartimento di Fisica ``Guglielmo 
Marconi'', Piazzale A. Moro 5, I--00185, Roma, Italy 
}

\date{Received 1999 September 16, Accepted 2000 January 6} 
\pubyear{2000} 

\begin{document} 
 
\maketitle 

\begin{abstract} 
We report the discovery of the optical counterpart 
of the 12.3\,s transient X--ray pulsar \src. 
We re--analysed archival \R PSPC observations of \src\, obtaining two new 
refined positions, $\sim$14\arcsec\ and $\sim$18\arcsec\ away from 
the previously published one, and a new spin period measurement. 
Based on these results we carried out optical and infra--red (IR) follow--up 
observations. Within the new error circles we found a relatively faint 
(V=20.1) early type reddened star (V--R=2.24). The optical spectrum  
shows a strong H$\alpha$ emission line. The IR observations 
of the field confirm the presence of an IR excess for the 
H$\alpha$--emitting star (K\arcmin=11.4, J--K\arcmin=1.94) which is likely 
surrounded by a conspicuous circumstellar envelope. 
Spectroscopic and photometric data indicate a B0--2 V--IIIe spectral--type 
star located at a distance of 3--5\,kpc and confirm the Be--star/X--ray 
binary nature of \src.       
\end{abstract} 


\section{Introduction} 
Be/X--ray binary systems (BeXBs) represent the majority of the known  
High Mass X--ray Binaries (HMXBs) hosting an accreting rotating magnetic 
neutron star 
(White, Nagase \& Parmar 1995). Phenomenologically, in the X--ray band, BeXBs can be 
divided into at least  
three subclasses: (i) bright transients which 
display giant X--ray outbursts up to L$_x$=10$^{38}$\,$\ergs$ (Type II; Stella 
et al. 1986) unrelated with the  orbital phase, with high spin--up rates, 
(ii) transients  which 
display periodic outbursts of relatively high luminosity 
(L$_x$\,$\simeq$\,10$^{36}$\,--\,10$^{37}$\,$\ergs$; Type I) generally occurring 
close to the periastron passage of the neutron star, and (iii)  
sources displaying no outbursts, but comparatively moderate variations 
(up to a factor of $\sim$ 10--100) and low--luminosity ($\leq$10$^{36}$ $\ergs$) 
pulsed persistent emission (Negueruela 1998). 
4U\,0115+634 (P=3.6\,s), V\,0332+30 (P=4.4\,s) 
and EXO\,2030+375 (P=41.7\,s) are all examples of the first two classes. Among the 
latter group, there are the well known X--ray pulsators 
X\,Per (P=835\,s) and RX\,J0146.9+6121 (P=1455\,s).
{\rc Moreover, differences between the Galaxy and the Magellanic Cloud population of 
BeXBs have been studied and are probably due to the different star formation rate  
(see Stevens et al. 1999)}.
B--emission (Be) spectral--type stars are characterized by high 
rotational velocities (up to 70\% of their break--up velocity), and 
by episodes of equatorial mass ejection which might produce a rotating ring of 
gas around the star at irregular time intervals. At optical wavelengths, 
Be stars are difficult to classify due to the presence of the 
circumstellar envelope responsible for the emission--lines.

The hard X--ray transient \src\ ($l_{II}$\,$\sim$\,262.0, $b_{II}$\,$\sim$\,--1.51) 
was  
discovered by the WATCH experiment onboard GRANAT in 1990 at a flux level of 
about 1 Crab in the 5--15 keV energy band (Sunyaev 1990). The source was later 
observed by GINGA (Makino 1990a, 1990b) and ROSAT as a part of the All Sky 
Survey (Hasinger \etal\ 1990) providing a much better positional accuracy 
(radius of 50\arcsec). Moreover pulsations at a period of 12.3 s were 
observed during the GINGA, ROSAT and ART--P observations (Makino 1990c; Aoki 
\etal\ 1992; Hasinger \etal\ 1990; Grebenev \& Sunyaev 1990). A target of 
opportunity ROSAT pointing performed in 1991 May allowed the 
source position to be determined with an uncertainty radius of 9\arcsec\ 
(Belloni \etal\ 1993). An optical follow--up observation of the stars within this 
error circle was performed at the ESO New Technology Telescope (NTT) 
in January 1991: no plausible optical counterpart was found down to a limiting 
magnitude of R$\sim$23.5 (Belloni \etal\ 1993). 

\src\ was also monitored by the Burst And Transient Source Experiment (BATSE) 
on the Compton Gamma Ray Observatory (CGRO) between 
April 1991 and July 1998. In particular seven outbursts were observed  
from April 1991 and June 1993 with a peak and intra--outburst flux 
of about 300\,mCrab and $<$10\,mCrab, respectively (Wilson et al. 1997). The  
recurrence time of 105--115 days was interpreted as the orbital period 
of the system. However, no further outbursts have been observed since July 1993 
either with CGRO/BATSE and the All Sky Monitor (ASM) on board the Rossi X--ray 
Timing Explorer.    
All these findings suggested that \src\ is a new Be--star/X--ray 
binary system in an eccentric orbit (Wilson et al. 1997). However the lack of 
any plausible optical counterpart remained a key point against this 
interpretation.

In this paper we report the discovery of the optical counterpart of 
\src\, a $V$=20.4 Be spectral--type star. Starting from three public ROSAT 
PSPC observations of \src\ we obtained two new independent position 
measurements (uncertainty radius of 10\arcsec). Within these new error circles 
we found a highly reddened star (V--R=2.24), the optical spectrum of which 
shows a strong H$\alpha$ emission line. The optical counterpart was 
also observed in the IR showing that this star is by 
far the brightest object in the field ($K\arcmin$=11.4). Our findings 
together, obtained from observations in three different energy bands,  
support the Be--star/X--ray binary nature of \src.

\section{X--ray observations} 
The PSPC (0.1--2.4 keV) detector on board \R observed the field including \src\ 
several times. In the \R public archive there are 5 observations performed 
between April and December 1991. PSPC images were accumulated in the 0.5--2.0 
keV range in order to reduce the strong and spatially inhomogeneous local 
background dominated by the soft X--ray emission from the Vela Supernova Remnant. 

Both a sliding cell and a Wavelet transform--based detection algorithm were used 
in order to characterize the physical parameters (position, count rate, S/N 
ratio, etc.) of \src\ 
when detected, and to obtain a 3$\sigma$ count rate upper limit in case of 
non--detection (Lazzati \etal\ 1999; Campana 
\etal\ 1999). Table\,1 summarizes the results of this analysis. 

\begin{table*}
\begin{flushleft}
\begin{minipage}{168mm}
\caption{\R and \A observations of \src.}
\begin{tabular}{llrccccl}
\hline \hline 
Pointing &  Instr.& Expos. & Start Time & Stop Time & Count rate$^{\dag}$ & 
Off--axis & Coordinates$^{\ddagger}$ \\
Number& & (s) &  & & (ct s$^{-1})$ & angle ($\arcmin$)&\\
\hline
  500015  &    PSPC &    1854 &  1991 Apr 25 13:05 &  1991 Apr 25 13:30 & $<$0.20 & 44& ---\\
  160062  &    PSPC &    2191 &  1991 May 05 16:41 &  1991 May 05 17:18 & 
  0.510$\pm$0.025 & 22& see Belloni \etal\ 1993\\
  160061  &    PSPC &    2005 &  1991 May 05 19:46 &  1991 May 05 20:21 &  
  0.067$\pm$0.018 & 0&R.A.=08$^h$\,35$^m$\,55$\fs$6 \\ 
          &         &         &                    &                    &         &    &
  Dec.=--43$\degr$\,11$\arcmin$\,07$\farcs$3   \\        
  500015  &    PSPC &    8666 &  1991 May 06 06:40 &  1991 May 06 14:58 & $<$0.11  & 44 & ---\\
  500128  &    PSPC &    1316 &  1991 Dic 04 01:19 &  1991 Dic 04 01:41 &  0.237$\pm$0.023 
  & 12 & ---\\
  500127--8 &  PSPC &    18235&  1991 Dic 17 11:31 &  1991 Dic 21 08:38 &  
  1.137$\pm$0.019 & 12 &R.A.=08$^h$\,35$^m$\,55$\fs$2 \\
          &         &         &                    &                    &         &    &
  Dec.= --43$\degr$\,11$\arcmin$\,10$\farcs$3  \\            
\hline 
  42023000&    GIS  &   35916 &  1994 Nov 28 07:45 &  1994 Nov 28 19:01 & $<$0.01 & 10& ---\\
\hline
\end{tabular}\\ 
\noindent $^{\dag}$ \R PSPC and \A GIS locally background corrected count 
rates were obtained in the 
0.5--2.0 keV and 0.5--10 keV energy bands, respectively. The \R PSPC count 
rates are vignetting and PSF corrected. Errors are 
at 1$\sigma$ level while upper limits at 3$\sigma$.\\
\noindent $^{\ddagger}$ Equinox 2000. Error radius of 10$\arcsec$ at 90\% 
confidence level.
\end{minipage}
\end{flushleft}
\end{table*}
\begin{figure}
\centerline{\psfig{figure=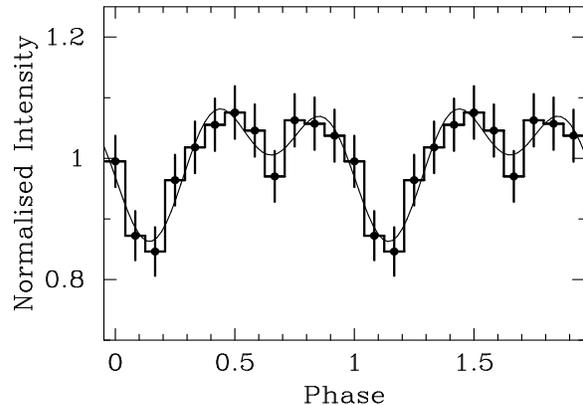,width=8cm,height=5cm} }
\caption{The 1991 December 17--21 \R PSPC  light curve folded at 
the best period  P=12.307\,s.}
\end{figure}
     
\src\ was detected in May and December 1991 observations as a highly variable 
source: from 0.067 
ct s$^{-1}$ (May 5) up to 1.14 ct s$^{-1}$ (December 21) corresponding to a 
variation of a factor of $\sim$20. We note that the flux we obtained for sequence 
160062 is consistent with that inferred by Belloni \etal\ (1993). 
The source was also detected during sequence 160061 (see Table\,1). 

For each \R observability window of \src\ we obtained an independent  
source position. These were determined to be R.A.=08$^h$\,35$^m$\,55$\fs$6,  
Dec.=--43$\degr$\,11$\arcmin$\,07$\farcs$3 (sequence 160061; May 91; equinox 2000), and 
R.A.=08$^h$\,35$^m$\,55$\fs$2, Dec.= --43$\degr$\,11$\arcmin$\,10$\farcs$3 
(sequence 500127--8; Dec 17--21; equinox 2000). 
The statistical uncertainty corresponds to an error radius of only 0$\farcs$5 
and 0$\farcs$2 for May 5 and December 17--21, respectively (90\% confidence 
level).  However due to the uncertainty  in the boresight correction the error radius 
increases to 
10$\arcsec$. We excluded from our position analysis sequence 160062 
as the source was close to the circular support rib (radius of $\sim$20$\arcmin$) 
of the PSPC, resulting in a higher uncertainty (error radius of 
$\ga$20$\arcsec$).  
We note that the new refined positions of \src\ lie about $\sim$18$\arcsec$ 
(160061) and $\sim$15$\arcsec$ 
(500127--8) away from that obtained from sequence 160062 (Belloni \etal\ 1993; 
error radius 9$\arcsec$). 

The \R event list and spectra of \src\ were extracted from a circle  
around the best X--ray position {\rc (with an extraction radius corresponding to an 
encircled energy of 90\% at the relevant off--axis angle)}. 
The photon arrival times were corrected to the barycentre of the 
solar system and a  background corrected 1\,s binned light curve accumulated 
for each observation. In order to maximize the period search sensitivity we 
analyzed only the observation with the highest statistics, namely sequence 
500127--8 merged. A power spectrum was calculated over the entire observation 
duration following the method described by Israel \& Stella (1996). 
To increase the search sensitivity we searched for significant peaks in a 
narrow period interval centered around the period value detected  by BATSE between 
1991 December 28 and 1992 January 2 (Wilson \etal\ 1997; 12.307\,s), thus only a few 
days after the \R observation. 
We found a significant peak at a confidence level of 99.5\% at a period of 
12.307$\pm$0.003\,s (90\% uncertainties are used through out this paper).
The pulsed fraction is 7\%$\pm$2\%, while the 0.1--2.4 keV pulse shape is 
well fitted by two sinusoidal waves (see Fig.\,1). A similar shape and pulsed 
fraction value were 
obtained at higher energies in 1991 December by BATSE (see Fig.\,5, 11 and 12 
in Wilson \etal\ 1997). 
\begin{table}
\begin{center}
\caption{\R PSPC spectral results for \src\  }
\begin{tabular}{llll}
\hline \hline 
Parameter &  May 1991 & May 1991 & Dec 1991  \\ 
          &  (160062) & (160061) & (500127--8) \\ 
\hline  
\multicolumn{4}{c}{Power--law model}\\
N$_H$ (10$^{22}$ cm$^{-2}$)& 2.3$\pm_{0.3}^{0.5}$ & 1.1$\pm_{0.4}^{0.8}$& 
                             2.3$\pm_{0.3}^{0.5}$ \\
$\Gamma$& 0.7\,(fixed)& 0.7\,(fixed) & 0.7$\pm0.7$ \\
F$_X$ (10$^{-11}$)& 1.73& 0.17& 5.46\\
L$_X$ (N$_H$=0; 10$^{35}$) & 2.58& 0.13 & 8.05\\
\hline 
\multicolumn{4}{c}{Blackbody model}\\
N$_H$ (10$^{22}$ cm$^{-2}$)& 2.0$\pm_{0.3}^{0.5}$ & 0.8$\pm_{0.4}^{0.8}$& 
                             2.0$\pm_{0.2}^{0.3}$ \\
kT (keV) & 1.0(fixed)& 1.0(fixed)& 1.0$\pm_{0.3}^{0.4}$\\ 
F$_X$ (10$^{-11}$)& 1.71& 0.16 & 5.44\\
L$_X$ (N$_H$=0; 10$^{35}$) & 1.781& 0.10& 5.50\\
Radius (km @ 5kpc) & 9.2& 2.1 & 16.7 \\
\hline
\end{tabular}
\end{center}
\noindent Note. --- The X--ray fluxes (units are erg\,cm$^{-2}$\,s$^{-1}$) 
and the unabsorbed luminosities (units are  erg\,s$^{-1}$; 
a distance of 5kpc was assumed) refer to the 0.1--2.4 keV energy band.
\end{table}

The PSPC Pulse Height Analyser (PHA) rates were grouped so as to contain a 
minimum of 20 photons per energy bin. The spectra of sequences 160062, 160061 
and 500127--8 (December 17--21) were fitted simultaneously. 
By using the brightest spectrum parameters as templates we fitted the three 
datasets keeping fixed the photon  index and the temperature of the 
power--law and blackbody, 
respectively, for the two fainter spectra. A power--law ($\Gamma$=0.7)  as well 
as a blackbody  (kT=1\,keV) model gave good fits ($\chi^2_{\nu}$=1.09 and  
$\chi^2_{\nu}$=1.11, respectively).

The field of \src\ was also observed on 1994 November 28 with the ASCA 
satellite. The source was not detected in the 0.5--10 keV band and a 
3$\sigma$ upper limit of 0.01 ct s$^{-1}$ was inferred (corresponding to 
$\sim$4$\times$10$^{-13}$erg s$^{-1}$ cm$^{-2}$ assuming the best power--law 
model parameters of Table\,2).

\section{Optical follow--up}

Based on the new X--ray positions, we carried out an optical follow--up from 
the ESO (La Silla, Chile) on 1999 January 18--20 with the 1.5\,m Danish 
telescope and on March 13 with the New Technology Telescope (NTT).  

Imaging and photometry in V, R and Gunn--i bands (200\,s exposure time each) 
were performed on 1999 January 18--19 with the Danish Faint Object Spectrometer
Camera (DFOSC), while spectroscopy of the brightest stars within the X--ray 
position uncertainty regions was obtained with the same instrument on 1999 
January 19--20. The data were reduced using standard ESO--MIDAS procedures for 
bias subtraction, flat--field correction, aperture photometry and one 
dimensional stellar and sky spectra extraction. Profile fitting photometry 
was also carried out with DAOPHOT\,II (Stetson 1987). Cosmic rays were 
removed from each frame, and the spectrum corrected for the atmospheric 
extinction and flux calibrated. 
\begin{figure*}
\centerline{\psfig{figure=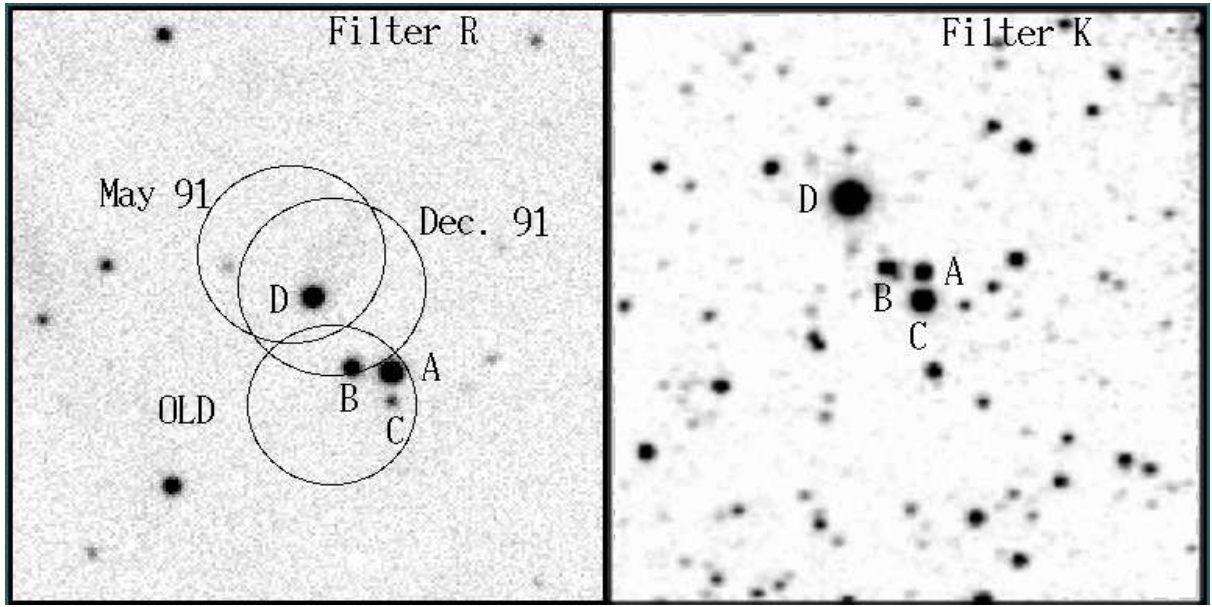,width=16cm,height=8cm} }
\caption{
R--band (left) and K\arcmin--band (right) images of the field of \src\ together with 
the new X--ray position uncertainty regions (10\arcsec\ radius) obtained with the 
1991 \R PSPC observations. North is top, east is left. Stars D represents the proposed optical 
counterpart of \src.} 
\end{figure*} 
Figure\,2 shows the field of \src\ in the R filter (left panel) with 
the new X--ray uncertainty circles obtained from the 1991 May and 
December \R observations superimposed. 
The earlier X--ray error circle (labelled as OLD) is also shown. 
Stars A and star C lie outside the new uncertainty regions ($\sim$2\arcsec\ and 
$\sim$6\arcsec\ away, respectively) while star D is the only object consistent 
with both circles and  it is located at R.A.\,=\,08$^h$\,35$^m$\,55$\fs$4 and 
Dec.\,=\,--43$\degr$\,11$\arcmin$\,11$\farcs$9 (equinox 2000; 
calibrated with DSS1 plates; uncertainty 1\arcsec). 
Photometry measurements for stars A, B, C and D are  
reported in Table\,3. 

Since the spectroscopic properties of star A were investigated by Belloni \etal\ 
1993 (star A is a late spectral type star without any 
significant emission features), we focused our attention on star D. 
We obtained three low--resolution (11\AA) spectra (30\,min exposure time 
each; 1\farcs5 slit; 1\farcs2 seeing) of star D on 1999 January 20 with a 
grism covering the 
spectral range 5200--10000\AA\ (see upper line of Fig.\,3) . Note that the blue part 
of the spectrum, being relatively faint (V=20.4), was below the instrument sensitivity 
(telescope+CCD+grism). After reduction, the spectra were summed to increase 
the S/N ratio. Star D was also observed on 1999 March 13 with the ESO 
Multi--Mode Instrument (EMMI) mounted on the adaptor--rotator at the Nasmyth B 
focus of the NTT. A 30\,min low--resolution (11\AA; 2\arcsec\ slit; 2\arcsec\ 
seeing) spectrum was obtained in the 4000--8500\AA\ range (see Fig.\,3; lower line). 

\begin{table} 
\begin{center} 
\caption{Optical and IR results for star {\rc A,B,C and D}} 
\begin{tabular}{llllllll} 
\hline \hline 
Star &  V & R & V--R & J & H & K\arcmin & J--K\arcmin \\ 
\hline 
{\rc  A }  & {\rc 18.9}  & {\rc 17.8} & {\rc 1.1} & {\rc 15.86} & {\rc 15.06} & {\rc 14.80} & {\rc 1.06}\\
{\rc   B }  & {\rc20.4}  & {\rc 19.0} & {\rc 1.4} & {\rc 16.20} & {\rc 15.24} & {\rc 14.98} & {\rc 1.22}\\
{\rc   C  } & {\rc 22.9}  & {\rc 20.8} & {\rc 2.1} & {\rc 15.40} & {\rc 13.81} & {\rc 13.29} &{\rc  2.11}\\
  D   & 20.4  & 18.2 & 2.2 & 13.33 & 12.26 & 11.39 & 1.94\\   
\hline  
\end{tabular} 
\end{center} 
Note --- Optical magnitude absolute uncertainty is $\sim$0.2, IR is $\sim$0.02.\\ 
\end{table} 
\begin{figure*} 
\centerline{\psfig{figure=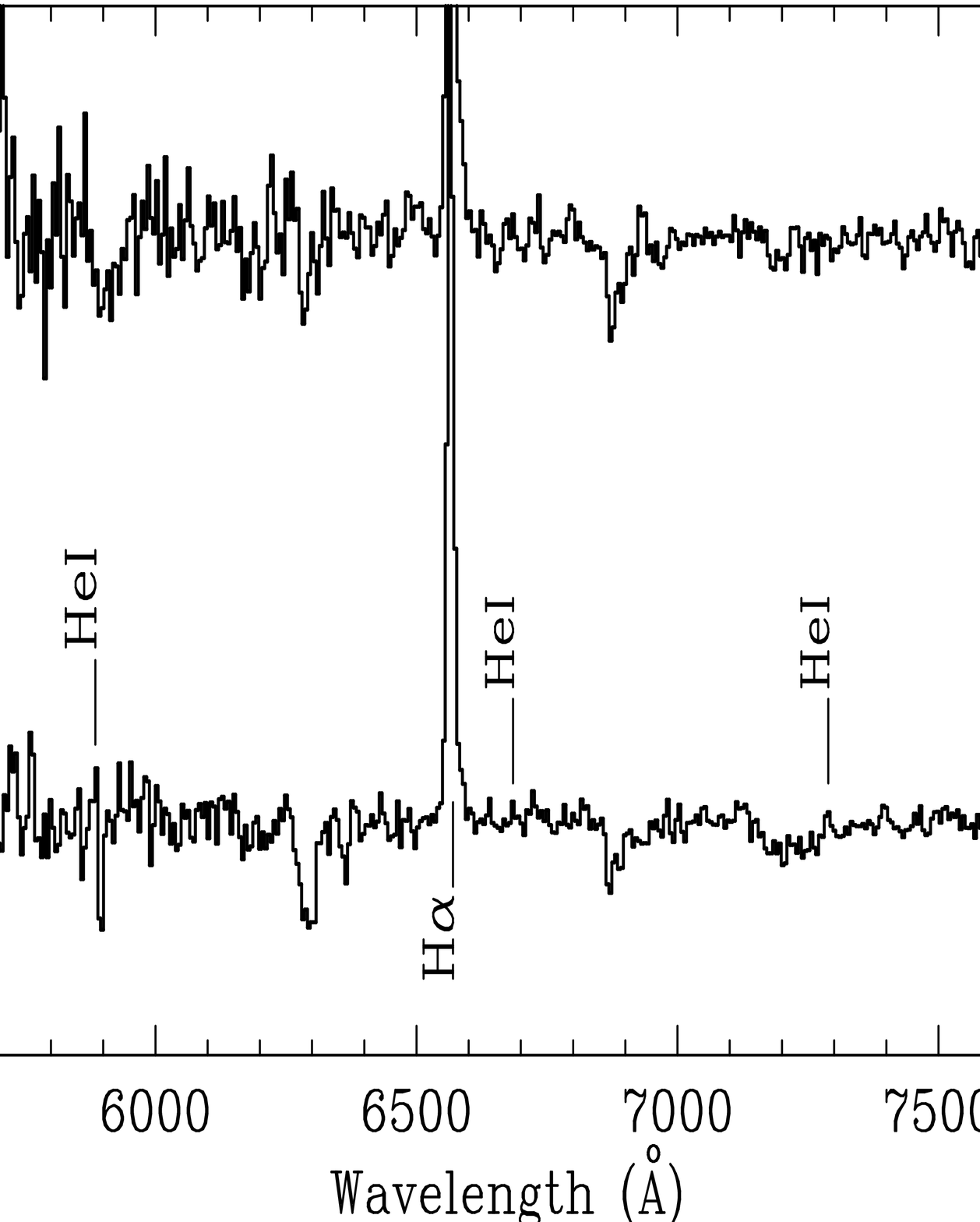,width=8cm,height=7cm} } 
\caption{Low--resolution spectra of \src\ obtained on 1999 January 20 
(upper line; Danish+DFOSC) and March 13 (lower line; NTT+EMMI). 
Fluxes are arbitrarly normalised and a constant was added to the January 20 
one.} 
\end{figure*} 

Due to its faintness, star D is difficult to study. 
The S/N ratio of the summed DFOSC spectrum is acceptable only above 
$\sim$6000\AA, while at $>$7500\AA\ the signal is 
dominated by interference fringes. 
The steep rise of the spectrum in the UV argues for a very hot and reddened object. 
The absence of strong forbidden lines rules out the possibility of a 
pre--main sequence object or a cataclysmic variable. 
Moreover the absence of strong absorption features points to an OB 
spectral--type star. The main features of the DFOSC spectrum are: (i) a 
strong H$\alpha$ emission--line with equivalent width EW$\sim$\,--30$\pm$2\,\AA, and 
(ii) pronounced interstellar absorption lines for Na\,II (5890\AA\ and 6270\AA). 
The H$\alpha$ line shows an extended and asymmetric profile. At 11\AA\ spectral 
resolution, this might indicate a line splitting, possibly due to the presence 
of a disk. All this evidence clearly points to the association of this object 
with the X--ray source. 

The EMMI spectrum has a slightly better S/N ratio and allows us to extend the 
spectral region into the 5500--8400\AA\ range. 
However, a precise spectral classification is not yet possible. 
A few Si\,II lines (e.g. the strong 5466.43--5466.87\AA\ doublet) 
are recognizable, suggesting a B2 spectral type. However the uncertainty 
on the spectral classification is large, ranging from an O8e to a B3e star. 
{\rc Emission features corresponding to He\,I lines (e.g. 6678 and 7281\,\AA) 
are also evident, while the He\,I lines at 7065 and 7818\AA\ look filled in}, 
as well as faint nebular [N\,II] lines on both 
sides of H$\alpha$, which most likely are due to the superposed emission from the  
Vela SNR (see Fig.\,3). 
In the EMMI spectrum the H$\alpha$ EW is --33$\pm$2\AA, {\rc while the FWZH is 
41\,$\pm$\,3\,\AA}, 
corresponding to a stellar wind terminal velocity V$_{wind}$\,$\simeq$1800 km s$^{-1}$. 
Deep interstellar lines and bands are clearly visible in the star spectra, 
the Na\,D2 doublet having an EW of 4.1$\pm$0.8\AA. 

{\rc Star B the position of which is still marginally consistent with the new X-ray 
error circle was observed on 1999 January 19. We obtained a low--resolution (11\AA) 
spectrum (1\,h exposure time; 2\farcs0 slit; 1\farcs7 seeing)  with a grism covering 
the 3500--7000\AA\ spectral range. Star B is even more difficult to study  being 
fainter than star D. 
Although the spectrum has a low S/N ratio, we detected large and deep absorption 
features typical of late type stars (probably an early M). Moreover no emission lines 
were visible in its spectrum.}

\section{IR observations} 
 
Infrared observations were carried out on 1993 April 1 during service observing 
time using the 
Infrared Imager Spectrograph (IRIS) instrument (Allen 1993) mounted at the Cassegrain 
focus of the 3.9m Anglo--Australian Telescope (AAT), Siding Springs Observatory, 
New South Wales. Images were obtained in the J (60\,s), H (20\,s) and 
K$^{\prime}$ (5 secs.) bands. The data were reduced using the Starlink Figaro 
software (Shortridge \etal\ 1998), and analysed using the PHOTOM package 
(Eaton \& Draper 1998). The standard stars HD 84090 and HD 100231 were used for 
calibration purposes, giving absolute uncertainties of around 0.02 magnitudes in 
each band (although the H band image was near the saturation limit of the detector 
and should be treated with a little more suspicion). 

{\rc Taking the colours of a B2V star from Wegner (1994) as (V--K)=--0.67,
(J--K)=--0.18 and (H--K)=--0.04, using the reddening relationships 
which gives E(J--K)/E(B--V)=+0.50, and combining these with the
observed values for \src, we find E(J--K)=+2.0 and thus E(B--V)=+4.0.
Taking the parameter R$_{ext}$=3.3 (R$_{ext}$=A$_V$/E(B--V)), we 
estimate an extinction of A$_V$=13.2 from the IR magnitudes.}

\section{Discussion} 
 
The X--ray, optical and IR observations of the field of \src's presented here, led 
to the identification of the optical counterpart of this 12\,s transient 
X--ray pulsar discovered in 1990. 
The measurement of the distance to \src\ based on the optical 
data is hampered by the uncertainties in the spectral classification. 

However some information can be inferred based on our optical and IR photometric 
measurements.  
The intrinsic $V-R$ color for \src\ is $\sim -0.1$ (assuming a main sequence 
star with spectral type in the O9--B3 range). 
Since the observed $V-R$ is $\sim$\,2.25 the reddening should 
amount to $E_{\rm V-R} \sim 2.35$, and assuming a standard reddening law 
(Fitzpatrick 1999) this converts to $A_R \sim 7$ (regardless of whether 
the reddening 
medium is uniformly distributed along the line of sight or is intrinsic 
to the source). Moreover from the X--ray spectral fits, we derive a 
$N_H$ of 1--2 $\times$ 10$^{22}$ which is in good agreement with the 
photometric reddening estimate (Bohlin et al. 1978). 
The observed magnitude of \src\ is $R \sim 18.2$ and the 
absolute $R$ magnitude of a star with the assumed spectral type is within 
$-2.5 and -5$. Taking into account the effects of reddening, a distance 
modulus within $13.7-16.2$ can be derived. This translates to a 
distance of $\sim$\,5-17\,kpc,  
well beyond the Vela SNR located at $\sim$\,500\,pc. 
However the Galaxy edge in the direction of \src\ is located at $\sim$6kpc, 
so we can put a limit on the spectral--type of star D assuming that it 
is placed at the Galactic border. For an O9, B0 or B2 dwarf star we obtain 
a reddened ($A_R \sim 7$) $R$ magnitude at 6 kpc (distance modulus of 
$\sim$ 14) of 16.5, 17 and 18.5, respectively. When compared with 
the observed R magnitude of 18.2 for star D, this suggests a B2Ve spectral--type 
star. 

Also the derived E(B--V)=+4.0 from the IR data suggests a highly reddened object.
Using the relationship for the contribution of the circumstellar (cs) environment
derived by Fabregat and Reglero (1990), where E$_{cs}$=0.0049--0.00185
EW(H$\alpha$), and given the observed EW(H$\alpha$)$\sim$--30\,\AA, we
estimate E$_{cs}$ to be only 0.06 magnitudes. This suggests that the bulk of
the extinction is interstellar rather than local to \src. {\rc This results 
confirm that the contribution of the disc (which is 
clearly present, given the large equivalent width inferred for H$\alpha$)  
to the observed IR colors is relatively small.} 

However the value of A$_V$$\sim$13 derived from the IR observations differs
significantly from that estimated from the optical data (where A$_R$$\sim$7, 
and thus A$_V$$\sim$9.4), suggesting that the object is either earlier
in spectral class than B2, closer that 5--6\,kpc, or possibly not a main
sequence object. The observed magnitudes are similar to those of EXO
2030+375 (see e.g. Negueruela 1998), which is compatible with a B0V star at
$\sim$3\,kpc or a B0III at $\sim$5\,kpc.

Based on both photometric and spectroscopic findings we conclude that star 
D is most likely a B0--2 V--IIIe star at a distance of 3--5\,kpc. 
A more accurate distance and spectral
classification will have to await detailed optical spectroscopic
observations in the blue end of the spectrum.

For  a distance of 5\,kpc and extrapolating the  0.1--2\,keV \R fluxes  
to the 1--10 keV band (we assume the spectral model fitted by Aoki \etal\ 1992 
using  Ginga data) we obtain an outburst peak luminosity 
L$_X$\,(1--10\,keV)\,$\simeq$\,5--8\,$\times$\,10$^{36}$\,erg~s$^{-1}$. A somewhat 
higher luminosity ($\simeq$10$^{37}$\,erg~s$^{-1}$) would be inferred 
by extrapolating the 20--100\,keV BATSE peak flux reported by Wilson et al. (1997) 
to the 1--10\,keV band. Such a peak luminosity is a typical value 
shown by X-ray pulsars which also display Type I outbursts 
(Stella et al. 1986; Negueruela 1998)  
occurring close to the time of periastron passage and with a periodic 
recurrence given by the orbital period of the system.


\section*{Acknowledgments} 
This work was partially supported through ASI grant. PR thanks the service 
observing programme of the AAO.

\vfill
\eject


\begin{thebibliography}{} 
\bibitem{ } Allen D., 1993, IRIS Users Manual, Anglo-Australian Observatory 
manual 30a, version 2.5 
\bibitem{ } Aoki T., Dotani T., Ebisawa K., et al., 1992, PASJ, 44, 641 
\bibitem{ } Belloni T., Hasinger G., Pietsch W., et al., 1993, A\&A, 271, 487
\bibitem{ } Bohlin R.C., Savage B.D., \& Drake J.F., 1978, ApJ, 224, 132
\bibitem{ } Campana S., Lazzati D., Panzera R., et al., 1999, ApJ, 524, in press
\bibitem{ } Eaton N. \& Draper P.W., 1998, PHOTOM Users guide version 1.7, Starlink User Note 45.8
\bibitem{ } Fabregat J. \& Reglero V., 1990, MNRAS, 247, 407
\bibitem{ } Fitzpatrick E.L., 1999, PASP, 111, 63
\bibitem{ } Grebenev S. \& Sunyaev R., 1991, IAU Circ. 5294
\bibitem{ } Hasinger G., Pietsch W., \& Belloni T., 1990, IAU Circ. 5142 
\bibitem{ } Lazzati D., Campana S., Rosati P., et al., 1999, ApJ, 524, in press
\bibitem{ } Makino F., 1990a, IAU Circ. 5142
\bibitem{ } Makino F., 1990b, IAU Circ. 5139
\bibitem{ } Makino F., 1990c, IAU Circ. 5148
\bibitem{ } Negueruela, I., 1998, A\&A, 338, 505
\bibitem{ } Shortridge K., Meyerdierks H., Currie M., et al., 1998, Figaro Users guide 
version 5.4-0, Starlink User Note 86.16
\bibitem{ } Stella L., White N. E., \& Rosner R., 1986, ApJ, 208, 669
\bibitem{ } Stetson P.B., 1987, PASP, 99, 191
\bibitem{ } {\rc Stevens J.B., Coe M.J., \& Buckley D.A.H., 1999, MNRAS, in press 
(astro--ph/9906106)}
\bibitem{ } Sunyaev R., 1990, IAU Circ. 5122
\bibitem{ } Wegner W., 1994, MNRAS, 270, 229
\bibitem{ } White, N. E., Nagase, F., \& Parmar, A. N., 1995 , in
{\it X--ray Binaries}, Eds. W.H.G. Lewin, J. van Paradijs \&
E.P.J. van den Heuvel (Cambridge University Press), p.1 .
\bibitem{ } Wilson C.A., Finger M.H., Harmon B.A., et al., 1997, ApJ, 479, 388 
\end{thebibliography}
\end{document}